\documentclass[prb,twocolumn,aps,floatfix]{revtex4}          

\usepackage{graphicx}
\usepackage{dcolumn}
\usepackage{bm}

\begin{document}

\preprint{}

\title{Excitons, biexcitons and trions in self-assembled (In,Ga)As/GaAs
  quantum dots: Recombination energies, polarization and radiative lifetimes
  versus dot height}

\author{Gustavo A. Narvaez}
\author{Gabriel Bester}
\author{Alex Zunger}

\affiliation{National Renewable Energy Laboratory, Golden, Colorado 80401, USA}

\date{\today}

\begin{abstract}
  We calculate the height dependence of recombination energies, polarization
  and radiative lifetimes of the optical transitions of various excitonic
  complexes: neutral excitons ($X^0$), negatively- ($X^{-}$) and
  positively-charged ($X^{+}$) trions, and biexcitons ($XX^0$) in lens-shaped,
  self assembled In$_{0.6}$Ga$_{0.4}$As/GaAs quantum dots. By using an
  atomistic pseudopotential method combined with the configuration-interaction
  method, we predict the following.
  (i) The recombination energy of the lowest transition of $X^-$ blue-shifts
  as height increases, whereas that of $X^+$ red-shifts. Remarkably, the
  recombination of $XX^0$ shows a red-shift at small heights, reaches a
  maximum shift, and then blue-shifts for taller dots. This feature results
  from the height dependence and relative magnitude of the inter-electronic
  direct Coulomb interaction.
  (ii) Changes in dot height lead to a bound-to-unbound crossover for $X^-$,
  $X^+$ and $XX^0$.
  (iii) When considering the $[110]$ and $[1\bar{1}0]$ directions, the lowest
  transitions of $X^0$ and $XX^0$ manifest $[110]$ vs $[1\bar{1}0]$ in-plane
  polarization anisotropy that switches sign as a function of height as well
  as alloy randomness. $X^-$ and $X^+$ show transitions with negligible
  polarization anisotropy regardless of height.
  (iv) The ground state of $X^0$ is split in a low-energy pair that is
  forbidden (dark) and a high-energy pair that is allowed; thus, at $T=0\;{\rm
    K}$ the radiative lifetime $\tau(X^0)$ is long ($\sim {\rm ms}$) due to
  the dark exciton. On the other hand, at $T=10\;{\rm K}$, $\tau(X^0)$
  decreases moderately as height increases and its magnitude ranges from
  $2$-$3\,{\rm ns}$. The ground state of $X^-$ and $X^+$, and that of $XX^0$
  is allowed (bright); so, $\tau(X^-)$, $\tau(X^+)$ and $\tau(XX^0)$ are fast
  ($\sim {\rm ns}$) even at $T=0\,{\rm K}$. These radiative lifetimes depend
  weakly on height. In addition, $\tau(X^-)\sim\tau(X^+)\simeq 1.1\,{\rm ns}$,
  while $\tau(XX^0)\simeq 0.5\,{\rm ns}$.
We compare our predictions with available spectroscopic data.  
\end{abstract}

\maketitle

%
%
\section{Introduction}

Single-dot spectroscopy makes it possible to probe dot-to-dot changes in the
excitonic properties of self-assembled InGaAs/GaAs quantum dots.
\cite{smith_PRL_2005,ware_PhysicaE_2005,skolnick_ARMR_2004,guffarth_PhysicaE_2004,rodt_papers,urbaszek_PRL_2003,finley_PRB_2001,findeis_SSC_2000,landin_papers} 
Both single-particle and many-particle aspects of these properties depend
non-trivially on the quantum dots size and shape,\cite{books+review}
reflecting not only simple quantum-confinement physics, but also electronic
structure effects such as interband, intervalley, spin-orbit and
strain-induced state coupling. The description of these effects require an
atomistic multi-band approach.
\cite{wang_PRB_1999,wang_APL_2000,zunger_pss.b_2001,TB_method,sheng_PRB_2005}
Here we adopt a method that is based on screened pseudopotentials and the
configuration-interaction approach, and address the changes with height of
recombination (emission) energies, polarization and radiative lifetimes of
various neutral and charged excitons: the neutral exciton ($X^0$), negatively-
($X^{-}$) and positively-charged ($X^{+}$) excitons, and the biexciton
($XX^0$) in lens-shaped, self-assembled alloyed In$_{0.6}$Ga$_{0.4}$As/GaAs 
quantum dots. Our predictions compare reasonably well with available spectroscopic data.  We
also compare our findings in In$_{0.6}$Ga$_{0.4}$As/GaAs dots with those in
pure, non-alloyed InAs/GaAs dots.

%
%
%
%
%
\section{Energetics of the monoexciton, charged trions and biexciton}

%
\begin{figure*}[ht]
\includegraphics[width=10.0cm]{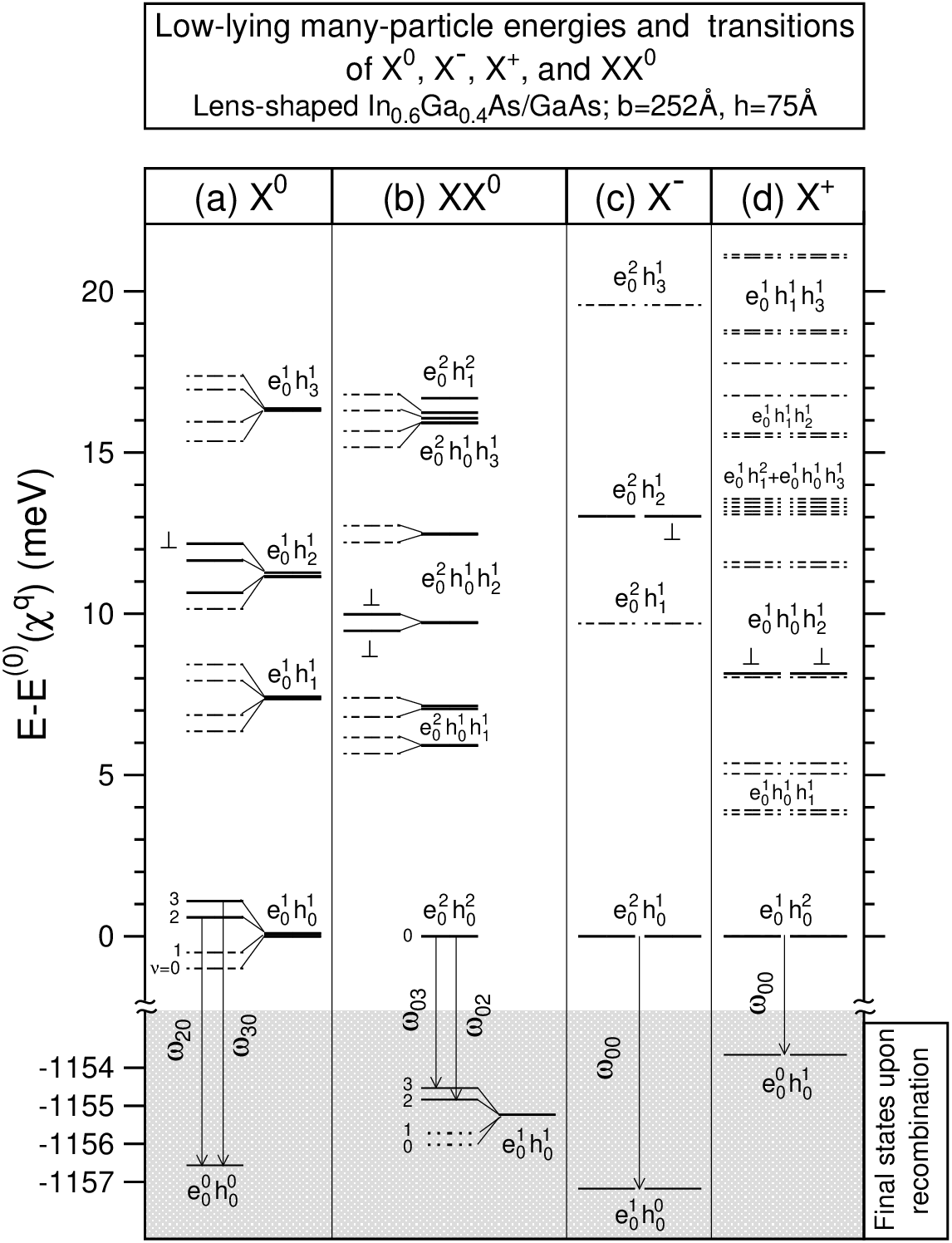}
\caption{{\label{Fig_1}} Quantitative many-particle, configuration-interaction
  low-lying energies and lowest optical transitions for (a) $X^0$, (b) $XX^0$,
  (c) $X^-$, and (d) $X^+$ in a lens-shaped In$_{\rm 0.6}$Ga$_{\rm
    0.4}$As/GaAs dot with base diameter $b=252\,${\AA} and height $h=75\,${\AA}. In
  each panel [(a)-(b)], the energy is shown relative to the ground-state
  energy of the corresponding exciton $\chi^q$ (=$X^0$, $X^-$, $X^+$, and
  $XX^0$). Two horizontally aligned dashes and dotted lines
  indicate, respectively, two-fold degenerate bright and dark levels. The
  dominant single-particle configuration of each excitonic level is indicated by
  $e^p_0h^{q}_jh^{q^{\prime}}_{j^{\prime}}$, where $p$, $q$ and $q^{\prime}$
  (=0, 1, 2) indicate, correspondingly, the occupation of the electron level
  $e_0$ and hole levels $h_j$ and $h_{j^{\prime}}$.  The energy of allowed
  transitions (vertical arrows) is indicated by $\omega_{if}$, where $i$ and
  $f$ indicate, respectively, the initial and final state upon recombination.
  Bright and dark excitonic levels are indicated, respectively, by solid and
  dotted lines. The fine-structure splittings are shown schematically.}
\end{figure*}

\subsection{Electronic structure of the excitonic manifolds}

We describe the basic electronic structure of the excitonic manifolds (Fig.
\ref{Fig_1}) before describing recombination processes. All excitonic states
are based on mixing and excitations of the single-particle states $\{h_0$,
$h_1$, $h_2$, $\cdots\}$ and $\{e_0$, $e_1$, $e_2$, $\cdots\}$, for holes and
electrons, respectively. These states are solutions to the single-particle
Schr\"odinger equation

\begin{equation}
\label{SP-effective_eq}
\left\{-\frac{1}{2}\nabla^2+V_{ext}({\bf R})+V_{scr}({\bf R})\right\}\psi_i={\cal E}_i\,\psi_i,
\end{equation}

\noindent where both the external (pseudo) potential $V_{ext}({\bf R})$ due to
the ion-ion and ion-electron interaction and the screening response to such
external potential $V_{scr}({\bf R})$ are expressed as a superposition of
screened atomic pseudopotentials

\begin{equation}
\label{pseudo}
V_{ext}({\bf R})+V_{scr}({\bf R})=V_{SO}+\sum_l\sum_{\alpha}v_{\alpha}[{\bf
  R}-{\bf R}^{(\alpha)}_l;{\rm Tr}(\widetilde\varepsilon)].
\end{equation}

\noindent Here, $V_{SO}$ is a non-local spin-orbit
interaction;\cite{williamson_PRB_2000} $v_{\alpha}$ is a screened
pseudopotential for atom of type $\alpha$ that depends on strain, and it has
been fitted to {\em bulk} properties of GaAs and InAs, including bulk band
structures, experimental deformation potentials and effective masses, as well
as local-density-approximation (LDA)-determined band
offsets.\cite{williamson_PRB_2000} The single-particle Shr\"odinger equation
[Eq. (\ref{SP-effective_eq})] includes not only quantum-confinement effects
(as in simple, one-band particle-in-a-box models), but also multi-band
coupling (light hole, heavy hole, conduction); inter-valley ($\Gamma$-$X$-$L$)
coupling; and spin-orbit coupling. Strain effects are present through the
relaxation, via a valence force field,\cite{williamson_PRB_2000} of the atomic
positions $\{{\bf R}^{(\alpha)}_l\}$ within the simulation supercell (quantum
dot+GaAs-matrix); and directly ``felt'' by the potential $v_{\alpha}[{\bf
  R}-{\bf R}^{(\alpha)}_l;{\rm Tr}(\widetilde\varepsilon)]$.  States $\{h_0$,
$h_1$, $h_2$,$\cdots\}$ and $\{e_0$, $e_1$, $e_2$,$\cdots\}$ form a basis for
the excitonic states.  We indicate the dominant single-particle configuration
of each excitonic level by
$e^p_ie^{p^{\prime}}_{i^{\prime}}h^{q}_jh^{q^{\prime}}_{j^{\prime}}$, where
$p$, $p^{\prime}$, $q$, and $q^{\prime}$ (=0, 1, 2) indicate, correspondingly,
the occupation of the electron levels $e_i$ and $e_{i^{\prime}}$, and hole
levels $h_j$ and $h_{j^{\prime}}$. Figure \ref{Fig_1} illustrates the
excitonic manifolds using a $75\,${\AA}-tall lens-shaped In$_{\rm
  0.6}$Ga$_{0.4}$As/GaAs quantum dot (base diameter $b=252\,${\AA}), as
obtained from configuration-interaction calculations based on the
pseudopotential single-particle description.

{\em Monoexciton.} The monoexciton $X^0$ has a ground-state
$|\Psi^{(0)}(X^0)\rangle$ created by occupying $h_0$ and $e_0$, denoted
$e^1_0h^1_0$ [Fig \ref{Fig_1}(a)].  However, configuration-interaction\cite{franceschetti_PRB_1999} (CI)
also mixes into $|\Psi^{(0)}(X^0)\rangle$ other states such as
$e^1_0h^1_1,e^1_0h^1_2,\cdots$. This is done by expanding the monoexciton
states $\{|\Psi^{(\nu)}(X^0)\rangle\}$ ($\nu=0,1,2,\cdots$) in a basis of
Slater determinants (configurations) $\{|\Phi(X^0)\rangle\}$ constructed in
the subspace of $N_e$ and $M_h$ electron and hole confined single-particle
levels, respectively:

\begin{equation}
\label{psi.nu.x0}
|\Psi^{(\nu)}(X^0)\rangle=\sum_{\kappa}C^{(\nu)}_{\kappa}(X^0)\,|\Phi_{\kappa}(X^0)\rangle, 
\end{equation}
  
\noindent where $C^{(\nu)}_{\kappa}(X^0)$ are the CI coefficients 
and $\kappa$ is a composite index that labels each Slater determinant.
\noindent Since $e_0$ and $h_0$ are each two-fold degenerate due to spin,
$e^1_0h^1_0$ has four-fold degeneracy at the single-particle level of Eqs.
(\ref{SP-effective_eq}) and (\ref{pseudo}). We next allow Coulomb 
electron-electron and hole-hole interactions. Direct Coulomb 
is given by $J^{\,(\mu\mu)}_{ij;ji}$ and exchange by $J^{\,(\mu\mu)}_{ij;ij}$, 
with the Coulomb scattering matrix elements given by 

\begin{widetext}
\begin{equation}
\label{J_eq}
J^{\,(\mu\mu^{\prime})}_{ij;kl}=\int\int {\rm d}{\bf R}{\rm d}{\bf
  R}^{\prime}\frac{\left[\psi^{(\mu)}_i({\bf
  R})\right]^{*}\left[\psi^{(\mu^{\prime})}_j({\bf
  R}^{\prime})\right]^{*}\left[\psi^{(\mu^{\prime})}_k({\bf
  R}^{\prime})\right]\left[\psi^{(\mu)}_l({\bf R})\right]}{\epsilon({\bf R},{\bf
  R}^{\prime})|{\bf R}-{\bf R}^{\prime}|}.   
\end{equation}
\end{widetext}

\noindent Here $\mu,\mu^{\prime}=e,h$; and $\epsilon({\bf R},{\bf R}^{\prime})$ 
is a microscopic, phenomenological dielectric constant.\cite{resta_PRB_1977} 
We also allow electron-hole direct Coulomb interaction $J^{\,(eh)}_{ij;ji}$ and
electron-hole exchange 

\begin{widetext}
\begin{equation}
\label{K_eq}
K^{\,(eh)}_{ij;kl}=\int\int {\rm d}{\bf R}{\rm d}{\bf
  R}^{\prime}\frac{\left[\psi^{(h)}_i({\bf
  R})\right]^{*}\left[\psi^{(e)}_j({\bf
  R}^{\prime})\right]^{*}\left[\psi^{(e)}_k({\bf R})\right]\left[\psi^{(h)}_l({\bf
  R}^{\prime})\right]}{\epsilon({\bf R},{\bf R}^{\prime})|{\bf R}-{\bf R}^{\prime}|}.
\end{equation}
\end{widetext}

\noindent Inclusion of these Coulomb interactions splits $e^1_0h^1_0$ into 
four distinct levels:\cite{bester_PRB_2003,bayer_PRB_2002} The lowest two are
spin-forbidden (``dark'') in the absence of spin-orbit coupling, and the
highest two are allowed (``bright''). The bright-dark splitting of the
ground-state levels shown in Fig. \ref{Fig_1}(a) is $84\,{\rm \mu eV}$. The
magnitude of this splitting increases up to $178\,\mu{\rm eV}$ for a
$20\,${\AA}-tall dot.  At $T\sim 0\,{\rm K}$ only the dark states are
populated, thus, the transition to the ground state $e^0_0h^0_0$ is
long-lived. Figure \ref{Fig_1}(a) (shaded area) shows transitions
$\omega_{20}$ and $\omega_{30}$ from the bright states to $e^0_0h^0_0$. The
low-lying excited states of the monoexciton correspond to excitations of the
hole, i.e. $e^1_0h^1_1$, $e^1_0h^1_2$ and $e^1_0h^1_3$, due to the much
smaller spacing of hole single-particle energy levels compared to that of the
electrons. Thus, the spacing of the excited states fingerprint the hole energy
level structure.  Figure \ref{Fig_1}(a) shows the first twelve excited states
that are derived from $e^1_0h^1_1$, $e^1_0h^1_2$ and $e^1_0h^1_3$. Each of
these states are four-fold degenerate and their fine structure is shown
schematically.  We find that three out of the four levels that arise from
$e^1_0h^1_2$ are optically allowed. In particular, one of the levels,
indicated as $\bot$, emits light that is polarized along $[001]$, while the
remaining two emit in-plane polarized light. These excited levels become
optically forbidden as the dot becomes flatter.

%
{\em Biexciton.} In contrast to the four levels comprising the monoexciton,
the biexciton $XX^0$ has a singly-degenerate, {\em closed-shell} ground state
($e^2_0h^2_0$) that is bright. Thus, even at $T\sim 0\,{\rm K}$ both emissions
$\omega_{03}$ and $\omega_{02}$ [Fig.  \ref{Fig_1}(b); shaded area] of the
biexciton are fast ($\sim {\rm ns}$). The biexciton has a non-trivial ladder
of excited states that is determined primarily by the relative magnitude of
the direct Coulomb hole-hole interaction and the single-particle energy
splittings of the hole states. The ladder begins with states derived from
$e^2_0h^1_0h^1_1$ at about $6\,{\rm meV}$ above the ground state and follows
with $e^2_0h^1_0h^1_2$ and $e^2_0h^1_0h^1_3$ at about $11\,{\rm meV}$ and
$16\,{\rm meV}$, respectively.  Due to the two-fold Kramers degeneracy of the
hole levels, $e^2_0h^1_0h^1_1$ is four-fold degenerate at the single-particle
level, these states split into four distinct states due to hole-hole and
electron-hole exchange [Fig \ref{Fig_1}(b)]. Similarly, the four states in
$e^2_0h^1_0h^1_2$ split in two groups of two. In this case, remarkably, the
splitting is about twice as big as that in $e^2_0h^1_0h^1_1$ and nearly five
times bigger than in $e^2_0h^1_0h^1_3$ ($\sim 500\,\mu{\rm eV}$).  Similarly
to the monoexciton case, the lowest split-off pair in $e^2_0h^1_0h^1_2$ is
optically active and emits light polarized along $[001]$ [$\bot$, Fig.
\ref{Fig_1}(b)].  Further, these states become darker as the dot becomes
flatter, being forbidden at $h=20\,${\AA}.  The subsequent excited state in
the ladder derives from the singly-degenerate closed-shell state $e^2_0h^2_1$
and is closely-spaced with $e^2_0h^1_0h^1_3$.  Note that albeit the splitting
between these states is small, the state derived from the closed-shell
configuration $e^2_0h^2_1$ is relatively ``inert'' in that it mixes weakly
with other states. In particular the weight of this configuration in the CI
expansion is $\sim 83\%$.  Below, we discuss the energy, polarization and
lifetime of $\omega_{03}$ and $\omega_{02}$.

%
{\em Trions.} The negatively- ($X^-$) and positively-charged ($X^+$) trions
have ground states that are bright, two-fold degenerate and arise from
occupying $e^2_0h^1_0$ and $e^1_0h^2_0$, respectively. Due to
configuration-interaction mixing, the ground states of $X^-$ and $X^+$ mix
with $e^2_0h^1_1$ and $e^1_0h^1_0h^1_1$, respectively.  As in the case of
$XX^0$, even at $T\sim 0\,{\rm K}$ the recombination $\omega_{00}$ of $X^-$
[Fig.  \ref{Fig_1}(c)] into state $e^1_0h^0_0$ and that of $X^+$ [Fig.
\ref{Fig_1}(d)] into state $e^0_0h^1_0$ is fast. For both trions, while there
are only two dipole-allowed transitions in the absence of spin-orbit coupling,
we predict four allowed transitions.  The first few excited states of $X^-$
correspond to occupying states derived primarily from $e^2_0h^1_1$,
$e^2_0h^1_2$ and $e^2_0h^1_3$ [Fig.  \ref{Fig_1}(c)].  (Naturally, these
states are mixed in with other states of $X^-$.) These excited states are
two-fold degenerate and lie, correspondingly, about $10\,{\rm meV}$, $13\,{\rm
  meV}$ and $20\,{\rm meV}$ above the ground state, as shown in Fig.
\ref{Fig_1}(c). Note the similarity of the low-lying excited states energy
spacing with that in $X^0$. In addition, note that one of the two states
derived from $e^2_0h^1_2$ is optically active and emits light polarized along
$[001]$ [$\bot$ in Fig.  \ref{Fig_1}(c)]. Again, these states become optically
forbidden for a $20\,${\AA}-tall dot.  The excited states of $X^+$ are more
complex and present a high density of states [Fig. \ref{Fig_1}(d)]; a
consequence of the closely spaced ($\sim 5\,{\rm meV}$) single-particle hole
levels: In contrast to $X^-$, all low-lying excited states show fine-structure
splitting due to electron-hole exchange.  The first excited states at about
$5\,{\rm meV}$ corresponds to occupying $e^1_0h^1_0h^1_1$, due to the two-fold
degeneracy of each of these single-particle states, there are eight excited
states that split in two groups of four, due to hole-hole exchange.  Occupying
$e^1_0h^1_0h^1_2$ leads to the next eight excited states around $10\,{\rm
  meV}$, which are also split in two quartets.  In this case, there is an
optically active (polarization $\Vert\,[001]$) pair of excited states that
belong to the lower-energy quartet [$\bot$, Fig. \ref{Fig_1}(d)]. As in the
other excitons, these states become dark as the dot height decreases.  The
next ten excited states about $14\,{\rm meV}$ are a mixture of eight
configurations derived from $e^1_0h^1_0h^1_3$ and two from $e^1_0h^2_1$.
Because $e_0$ is {\em half-filled}, $e^1_0h^2_1$ is very ``reactive'' in the
sense that it heavily mixes via configuration interaction. Note that this is
significantly different from the case of configuration $e^2_0h^2_0$ in the
biexciton [Fig.  \ref{Fig_1}(b)].  Higher excited states correspond to
occupying $e^1_0h^1_1h^1_2$ and $e^1_0h^1_1h^1_3$. Below, we discuss the
energy, as well as polarization and lifetime of $\omega_{00}$ for $X^-$ and
$X^+$.

%
%
%
%
\subsection{Recombination energies}
\label{recomb.sec}

We define the recombination energy upon recombination of an electron-hole pair
in $\chi^q$ as the difference between the total energies of the initial state
$|\Psi^{(i)}(\chi^q)\rangle$ and the final state
$|\Psi^{(f)}(\chi^q-1)\rangle$. Namely,

\begin{equation}
\label{omega_if.eq}
{
\begin{array}{rcl}
\omega_{i0}(X^0)&=&E^{(i)}(X^0) \\
\omega_{if}(X^-)&=&E^{(i)}(X^-)-{\cal E}^{\,(e)}_f \\
\omega_{if}(X^+)&=&E^{(i)}(X^+)+{\cal E}^{\,(h)}_f \\
\omega_{if}(XX^0)&=&E^{(i)}(XX^0)-E^{(f)}(X^0). 
\end{array}
}
\end{equation}

\noindent where $i$ and $f$ label an initial and final state,
respectively; ${\cal E}^{\,(e)}_f$ and ${\cal E}^{\,(h)}_f$ are, respectively,
the single-particle energy of the electron and hole in the final state; and
$E^{(\nu)}(\chi^q)$ is the multi-particle, configuration-interaction energy of
state $|\Psi^{(\nu)}(\chi^q)\rangle$.  [In Eq. (\ref{omega_if.eq}), the energy
of the ground state of the system in the absence of excited electron-hole
pairs is taken to be zero.]  Optical experiments like photoluminescence probe
electron-hole recombination transitions that are allowed
(bright).\cite{cardona_book}
Figure \ref{Fig_1}(a)-(d) indicates with arrows the lowest bright
recombination transitions of exciton $\chi^q$, for a $75\,${\AA}-tall
lens-shaped In$_{\rm 0.6}$Ga$_{\rm 0.4}$As/GaAs dot with base diameter
$b=252\,${\AA}. The final states upon electron-hole recombination are
presented in the shaded area of Fig.  \ref{Fig_1}.  Figure \ref{Fig_2} shows
the recombination energies $\omega_{00}(X^-)$, $\omega_{00}(X^+)$ and
$\omega_{03}(XX^0)$ calculated at the many-particle, configuration-interaction
level as a function of dot height. These recombination energies are shown
relative to the lowest recombination energy $\omega_{20}(X^0)$ of the
monoexciton (the latter energy is shown in the top axis of Fig.  \ref{Fig_2}).
Thus, the results correspond to {\em spectroscopic shifts} that are currently
measured by several groups. Prominent features:

%
\begin{figure}[htb]
\includegraphics[width=8.5cm]{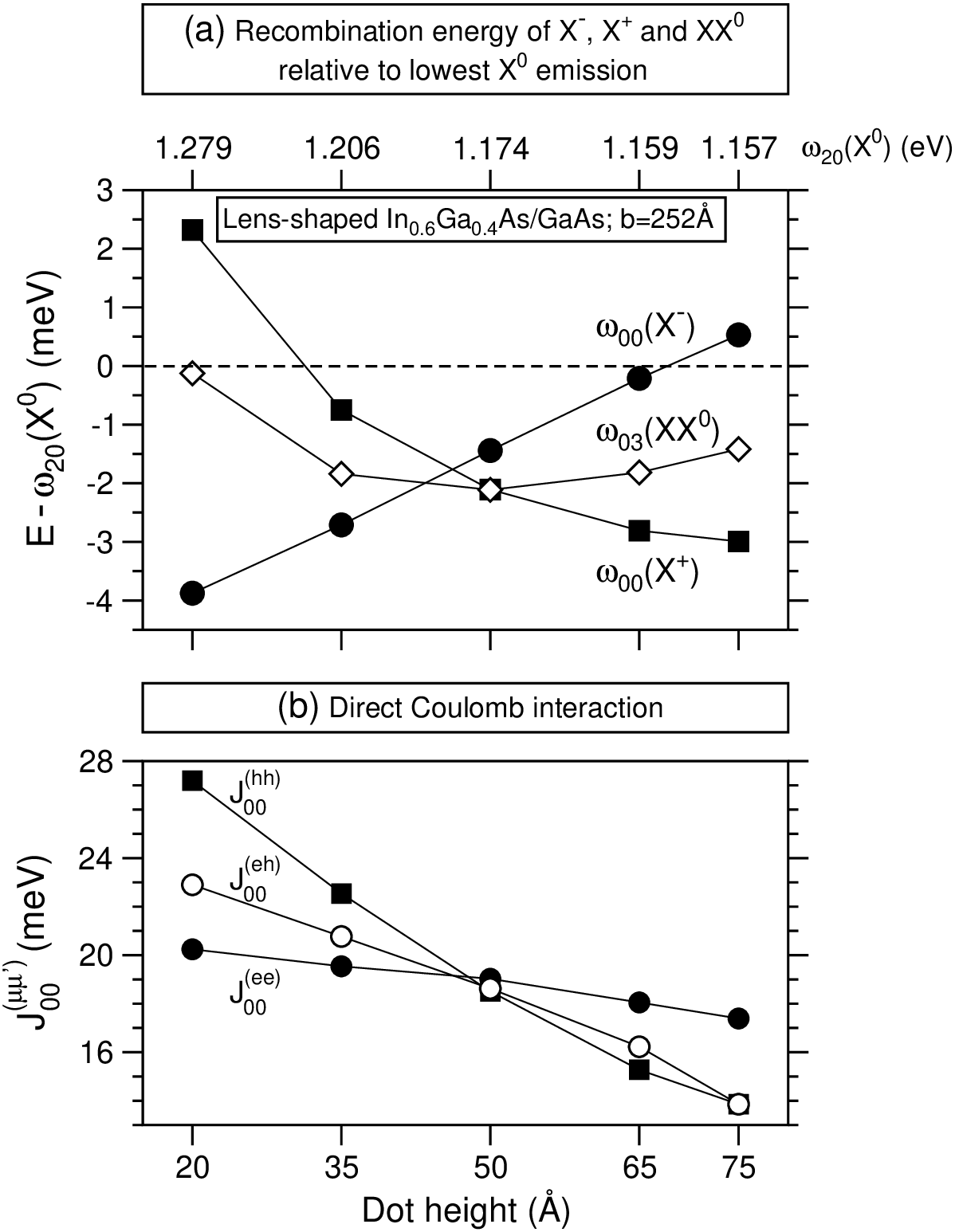}
\caption{{\label{Fig_2}} (a) Recombination energies $\omega_{00}(X^-)$ and
  $\omega_{00}(X^+)$, and $\omega_{03}(XX^0)$ (see text and Fig. \ref{Fig_1})
  in In$_{\rm 0.6}$Ga$_{\rm 0.4}$As/GaAs quantum dots (base $b=252\,$\AA) with
  different heights. The energy is shown as a spectroscopic shift relative to
  $\omega_{20}(X^0)$ [see Fig. \ref{Fig_1}(a)]. (b) Screened direct Coulomb
  interaction $J^{\,(\mu\mu^{\prime})}_{00}$ versus dot height. At
  $h=50\,${\AA} a nearly ``symmetric'' regime [$J^{\,(eh)}_{00}\sim
  J^{\,(hh)}_{00}\sim J^{\,(ee)}_{00}$] is attained.}
\end{figure}

(i) The recombination energy $\omega_{00}(X^{-})$ blue-shifts as height
increases; in contrast, $\omega_{00}(X^{+})$ red-shifts. These trends have
been explained by Bester and Zunger in Ref. \onlinecite{bester_PRB_2003b} by
adopting the Hartree-Fock approximation.

(ii) For the flattest dot ($h=20\;${\AA}) the ordering
$\omega_{00}(X^+)>\omega_{20}(X^0)>\omega_{00}(X^-)$ of the emission energies
and the relative magnitude of the spectroscopic shifts
$\omega_{00}(X^+)-\omega_{20}(X^0)<\omega_{20}(X^0)-\omega_{00}(X^-)$ agree
with photoluminescence (PL) data.
\cite{ware_PhysicaE_2005,skolnick_ARMR_2004,guffarth_PhysicaE_2004,finley_PRB_2001}
These relationships are explained at the Hartree-Fock level of approximation,
as in Ref. \onlinecite{bester_PRB_2003b}, in which one derives

\begin{eqnarray}
\omega_{20}(X^0)&=&[{\cal E}^{\,(e)}_0-{\cal E}^{\,(h)}_0]-J^{\,(eh)}_{00},\\
\omega_{00}(X^+)&=&[{\cal E}^{\,(e)}_0-{\cal E}^{\,(h)}_0]+J^{\,(hh)}_{00}-2\,J^{\,(eh)}_{00},\\
\omega_{00}(X^-)&=&[{\cal E}^{\,(e)}_0-{\cal E}^{\,(h)}_0]+J^{\,(ee)}_{00}-2\,J^{\,(eh)}_{00}.
\end{eqnarray}

\noindent Then, the relationship $J^{\,(hh)}_{00}>J^{\,(eh)}_{00}>J^{\,(ee)}_{00}$ that
holds at $h=20\,${\AA} [Fig. \ref{Fig_2}(b)] reveals the ordering of the
emission energies and the magnitude of the spectroscopic shifts.
Regarding the magnitude of the spectroscopic shifts,
$\omega_{00}(X^+)-\omega_{20}(X^0)=2.5\;{\rm meV}$ and
$\omega_{00}(X^-)-\omega_{20}(X^0)=-3.7\;{\rm meV}$ agree reasonably well with
(a) $2\;{\rm meV}$ and $-6\;{\rm meV}$, respectively, that Ware and coworkers
have recently observed in an InAs/GaAs dot (size unspecified) with monoexciton
emission at $\sim 1.268\;{\rm eV}$ (Ref. \onlinecite{ware_PhysicaE_2005});
also with (b) $\omega_{00}(X^-)-\omega_{20}(X^0)=-5\;{\rm meV}$ measured by
Smith {\em et al.} in an (In,Ga)As/GaAs dots (size unspecified) with emission
energy at $\sim 1.319\;{\rm eV}$ (Ref. \onlinecite{smith_PRL_2005}); and with
(c) the value of $-5.8\;{\rm meV}$ for $\omega_{00}(X^-)-\omega_{20}(X^0)$
observed by Finley and coworkers in an (In,Ga)As/GaAs dots [$b=(230\pm
70)\;${\AA}, $h=(25\pm 10)\;${\AA}] with emission at $\sim 1.263\;{\rm eV}$
(Ref.  \onlinecite{finley_PRB_2001}).

(iii) As height increases, $\omega_{03}(XX^0)$ red-shifts at small heights,
reaches a maximum shift of nearly $-2\;{\rm meV}$ at $h\sim 50\;${\AA}, and
then it moderately blue-shifts for taller dots. In addition, at $h=50\;${\AA}
the emission energy of $XX^0$ coincides with that of $X^{+}$. As in (ii),
these results are explained at the Hartree-Fock level, which predicts

\begin{eqnarray}
\label{wxx0}
\omega_{03}(XX^0)-\omega_{20}(X^0)&=&J^{\,(ee)}_{00}+J^{\,(hh)}_{00}-2\,J^{\,(eh)}_{00},
\\
\label{wx+}
\omega_{03}(XX^0)-\omega_{00}(X^+)&=&J^{\,(ee)}_{00}-J^{\,(eh)}_{00}.
\end{eqnarray}   

\noindent Here, we have neglected the small ($\sim 1$-$6\;\mu{\rm eV}$) splitting of
the monoexciton bright states. Then, by analyzing the height dependence of the
direct Coulomb interactions $J^{\,(ee)}_{00}$, $J^{\,(hh)}_{00}$ and
$J^{\,(eh)}_{00}$ [Fig. \ref{Fig_2}(b)] we find that Eq. (\ref{wxx0}) predicts
the observed height dependence of $\omega_{03}(XX^0)$, although the actual
magnitude of the emission is not quantitatively predicted due to
correlations.\cite{shumway_PRB_2001,narvaez_PRB_tobepublished} In addition,
Eq. (\ref{wx+}) reveals the coincidence of the emission of $XX^0$ and that of
$X^+$ as a result of balancing the magnitudes of $J^{\,(ee)}_{00}$ and
$J^{\,(eh)}_{00}$ at $h=50\,${\AA}. The latter balance arises from the similar
degree of localization of $\psi^{(e)}_0$ and $\psi^{(h)}_0$ within the
dot.\cite{narvaez_condmat}

(iv) $\omega_{03}(XX^0)-\omega_{20}(X^0)=-2.0\;{\rm meV}$ at $h=50\;${\AA}
agrees with the value of $-2.0\pm 0.1\;{\rm meV}$ for $XX^0$ measured by
Finley and coworkers in PL experiments in an (In,Ga)As/GaAs dot [$b=(230\pm
70)\;${\AA} and $h=(25\pm 10)\;${\AA}] with exciton ground-state emission at
$\sim 1.345\;{\rm eV}$ (Ref.  \onlinecite{finley_PRB_2001}). In addition, this
result of $-2.0\;{\rm meV}$ for the biexciton shift agrees remarkably well
with the value of $-2\;{\rm meV}$ measured by Rodt {\em et al} in InAs/GaAs
dots ($b=100-200\;${\AA}, height unspecified) with monoexciton emission
energies ranging between $1.260\;{\rm eV}$ and $1.295\;{\rm eV}$ (Ref.
\onlinecite{rodt_papers}); and also with the measured shift of $-2.3\;{\rm
  meV}$ observed by Urbaszek and coworkers in (In,Ga)As/GaAs dots (size
unspecified) with monoexciton emission energy of $1.294\;{\rm eV}$ (Ref.
\onlinecite{urbaszek_PRL_2003}).
The value of about $-1.7\,{\rm meV}$ for $\omega_{03}(XX^0)-\omega_{20}(X^0)$
that we find at $h=35\,${\AA} and $65\,${\AA} agrees well with the value of
$-1.6\,{\rm meV}$ observed by Bayer and coworkers in (In,Ga)As/GaAs dots
[$b=(500\pm 30)\;${\AA}] with exciton emission at $1.428\,{\rm eV}$ (Ref.
\onlinecite{bayer_PRB_1998}). Our results also agree satisfactorily with the
value of $-2.7\;{\rm meV}$ measured by Findeis and coworkers in dots (size
unspecified) with monoexciton emission energy at $1.284\;{\rm eV}$ (Ref.
\onlinecite{findeis_SSC_2000}).

%
%
%
%
%
\subsection{Binding energies}
\label{binding.mp}

The binding energies of excitons $\chi^q$ are defined as

\begin{equation}
\label{binding.eqs_CI}
{
\begin{array}{rcl}
\Delta(X^0)&=&\big[{\cal E}^{\,(e)}_0-{\cal E}^{\,(h)}_0\big]-E^{(0)}(X^0) \\
\Delta(X^-)&=&\big[{\cal E}^{\,(e)}_0+E^{(0)}(X^0)\big]-E^{(0)}(X^-) \\
\Delta(X^+)&=&\big[-{\cal E}^{\,(h)}_0+E^{(0)}(X^0)\big]-E^{(0)}(X^+) \\
\Delta(XX^0)&=&2E^{(0)}(X^0)-E^{(0)}(XX^0). 
\end{array}
}
\end{equation}

\noindent Exciton $\chi^q$ is said to be {\em bound} when the binding energy
$\Delta(\chi^q)$ is positive. Conversely, $\Delta(\chi^q)<0$ implies the
exciton is unbound. Note that the binding energy is defined with respect to
the ground-state energy of {\em dissociated} excitonic complexes.  For
instance, the binding energy of $X^0$ is defined with respect to the energy of
a non-interacting electron-hole pair. In turn, the binding energy of the
biexciton $XX^0$ is defined with respect to the total energy of two
non-interacting monoexcitons [Eq. (\ref{binding.eqs_CI})].

%
\begin{figure}[htb]
\includegraphics[width=8.5cm]{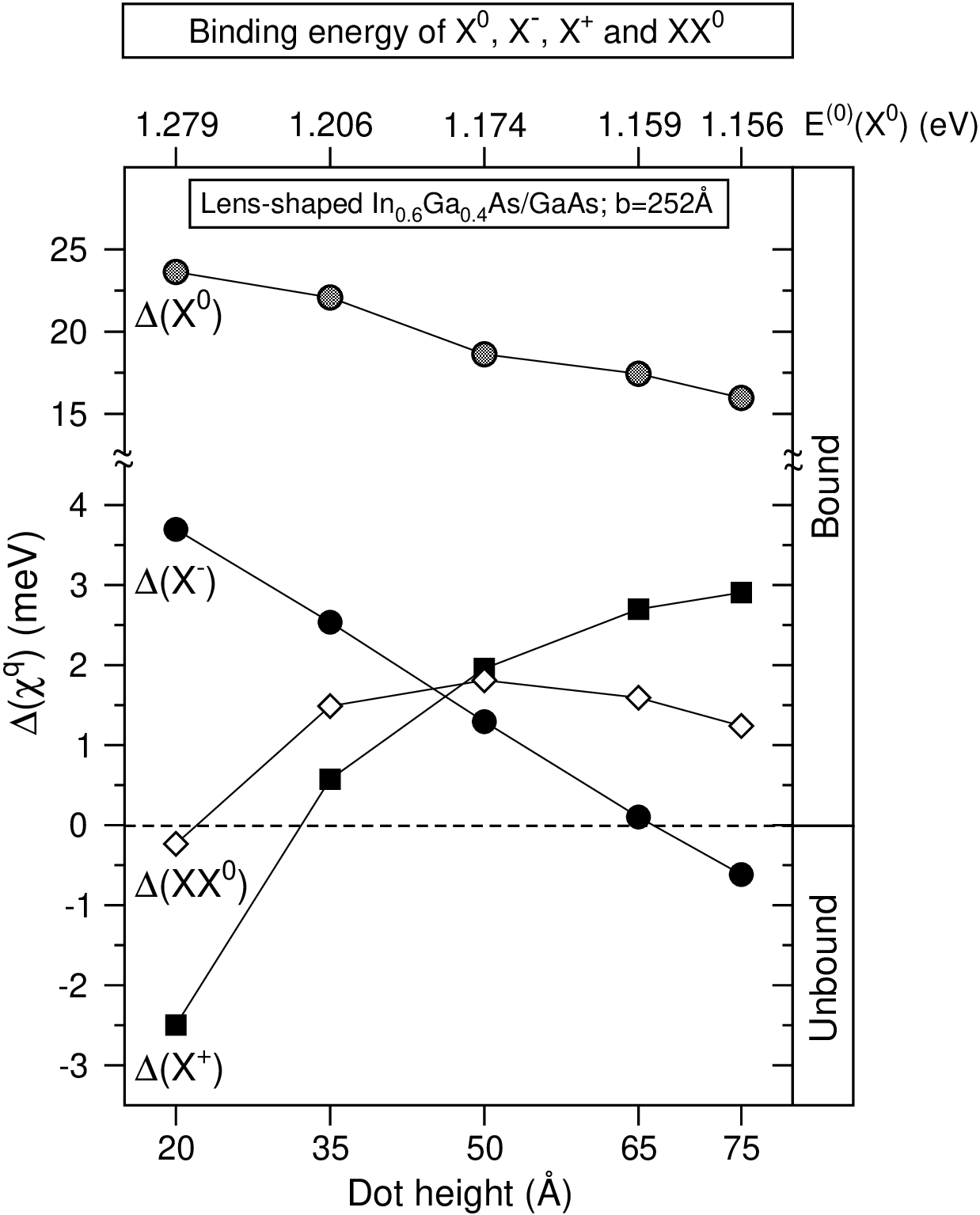}
\caption{{\label{Fig_3.mp}} Binding energy $\Delta(\chi^q)$
  [Eq. (\ref{binding.eqs_CI})] of $X^0$, $X^-$ and $X^+$, and $XX^0$ as a
  function of dot height. The ground-state energy $E^{(0)}(X^0)$ of $X^0$ is
  also shown (top axis). For $X^-$, $X^+$, and $XX^0$, height drives a
  bound-to-unbound crossover.}
\end{figure}

Figure \ref{Fig_3.mp} shows $\Delta(X^0)$, $\Delta(X^-)$, $\Delta(X^+)$ and
$\Delta(XX^0)$ as well as $E^{(0)}(X^0)$ as a function of dot height. 
$\Delta(X^0)$ decreases with increasing height, and it is well approximated by
$J^{eh}_{00}$ as correlation effects are relatively small ($\sim 2\;{\rm
  meV}$). The height dependence of the binding energy of $X^-$, $X^+$, and
$XX^0$ follows the height dependence of the spectroscopic shifts shown in Fig.
\ref{Fig_2}(a). This is so because the bright-dark splitting for the
monoexciton $X^0$ is small $\sim 80$-$180\;{\rm \mu eV}$. As expected from the
spectroscopic shifts results [Fig.  \ref{Fig_2}(a)], the height dependence of
the binding energy is qualitatively different for each exciton $\chi^q$.
(i) $\Delta(X^-)$ is bound ($\sim 4\;{\rm meV}$) for the
flattest dot ($h=20\;${\AA}) and it decreases almost linearly, becoming
unbound for dots taller than $h=65\;${\AA}.
(ii) In contrast, $\Delta({X^+})$ is unbound for the flattest dot and
increases up to $3\;{\rm meV}$, becoming bound slightly below $h=35\;${\AA}.
(iii) $\Delta(XX^0)$ does not depend monotonically on the
gap, reaching a maximum around $h=50\;${\AA}. 
In addition, $XX^0$ is bound [$\Delta(XX^0)>0$] for all heights above
$20\;${\AA} while unbound for the flattest dot ($h=20\;${\AA}). The latter is
due to an interplay between Hartree-Fock and correlation contributions to
binding, $\Delta_{\rm HF}(XX^0)$ and $\delta(XX^0)$, respectively, that
results in correlation being insufficient to bind $XX^0$. Namely,

\begin{eqnarray}
\label{Correlation}
\Delta(XX^0)&=&\Delta_{\rm HF}(XX^0)+\delta(XX^0) \\
& = & \{2J^{\,(eh)}_{00}-[J^{\,(ee)}_{00}+J^{\,(hh)}_{00}]\}+\delta(XX^0) \nonumber \\
 &=& -1.6\;{\rm meV}+1.4\;{\rm meV} \nonumber \\ 
&=& -0.2\;{\rm meV}, \nonumber
\end{eqnarray}

\noindent It should be noted that Rodt and coworkers\cite{rodt_papers}
observed (in photoluminescence) a bound-unbound crossover for $XX^0$ as the
monoexciton emission energy of InGaAs/GaAs quantum dots decreased. Those
authors also calculated the binding energy of $XX^0$, using an 8-band ${\bf
  k}\cdot{\bf p}$ model, and suggested that the reduction of correlation
effects was responsible for unbinding $XX^0$ as the gap
increased.\cite{rodt_papers}

%
%
%
%
%
\section{Polarization anisotropy of optical transitions}

When an electron-hole pair in exciton $\chi^q$ recombines optically, the
transition is characterized by both the transition energy
$\omega_{if}(\chi^q)$ [Eq.  (\ref{omega_if.eq})] and the transition dipole
matrix element

\begin{equation}
\label{dipole.element}
{M}^{(\hat{\bf e})}_{if}(\chi^q)=\langle\Psi^{(f)}(\chi^q-1)|\hat{\bf e}\cdot{\bf
  p}|\Psi^{(i)}(\chi^q)\rangle . 
\end{equation}

\noindent Here, ${\bf p}$ is the electron momentum and $\hat{\bf e}$ is
the polarization vector of the electromagnetic field.\cite{cardona_book}

%
The dipole matrix elements ${\bf M}^{(\hat{\bf e})}_{if}(\chi^q)$ [Eq.
(\ref{dipole.element})] depend on the polarization vector $\hat{\bf e}$; so,
it is natural to quantify what is the degree of polarization anisotropy
between different polarizations $\hat{\bf e_{1}}$ and $\hat{\bf e_{2}}$.
Therefore, we introduce the recombination (emission or pholuminescence)
intensity spectrum of exciton $\chi^q$ for polarization $\hat{\bf e}$,

\begin{equation}
I^{(\hat{\bf e})}(\omega,T;\chi^q)=\sum_{i,f}\big|{
    M}^{(\hat{\bf
    e})}_{if}(\chi^q)\big|^2\,P_{i}(T;\chi^q)\,\delta[\omega-\omega_{if}(\chi^q)].
\end{equation} 

\noindent Here, 

\begin{equation}
\label{bolzmann.n}
P_i(T;\chi^q)={\cal N}\exp\{-[E^{(i)}(\chi^q)-E^{(0)}(\chi^q)]/k_B\,T\}
\end{equation}

\noindent is the occupation (Boltzmann) probability of the 
initial state $|\Psi^{(i)}(\chi^q)\rangle$ at temperature $T$; ${\cal N}$ is a
normalization constant such that $\sum_{i}P_i(T;\chi^q)=1$ and $k_B$ is the
Boltzmann constant. Then, as {\em in-plane} polarizations $\hat{\bf
  e}_1=[110]$ and $\hat{\bf e}_2=[1\bar 10]$ have been probed extensively, we
introduce the in-plane polarization anisotropy parameter $\lambda$,

\begin{equation}
\label{lambda_eq}
\lambda(\omega,T;\chi^q)=\frac{I^{([110])}(\omega,T;\chi^q)-I^{([1\bar 10])}(\omega,T;\chi^q)}{I^{([110])}(\omega,T;\chi^q)+I^{([1\bar 10])}(\omega,T;\chi^q)}.
\end{equation}

\noindent Thus, $\lambda=1$ indicates an optical
transition that is {\em fully} polarized along the $[110]$ direction, while 
$\lambda=-1$ indicates one {\em fully} polarized along $[1\bar 10]$.

%
\begin{figure}[bt]
\includegraphics[width=8.5cm]{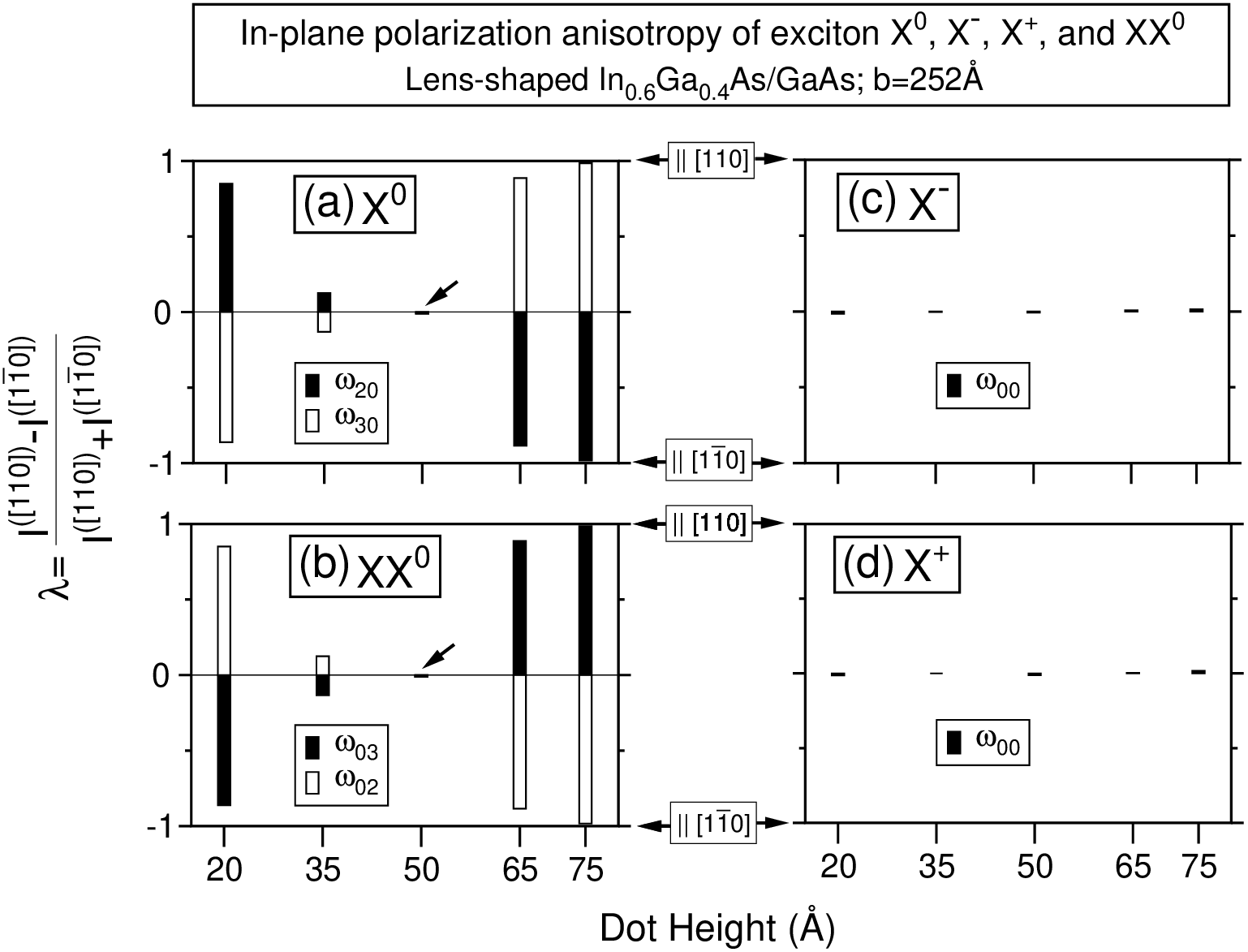}
\caption{{\label{Fig_4}}In-plane polarization anistropy $\lambda$ 
  [Eq. (\ref{lambda_eq})] for the lowest optical transitions of (a) $X^0$, (b)
  $XX^0$, (c) $X^-$, and (d) $X^+$. (See Fig. \ref{Fig_1}.) The transitions of
  $X^0$ and $XX^0$ are polarized along $[110]$ and $[1\bar 10]$, and the
  degree of this polarization strongly depends on height. For $X^-$ and $X^+$,
  the transitions are not polarized neither along $[110]$ nor $[1\bar 10]$.}
\end{figure}

\subsection{In-plane polarization anisotropy of the lowest optical transitions 
of $X^0$, $X^-$, $X^+$, and $XX^0$}

%
Figure \ref{Fig_4} shows the in-plane polarization anisotropy $\lambda$ for
the lowest optical transitions of (a) $X^0$, (b) $XX^0$, (c) $X^-$, and (d)
$X^+$ as a function of dot height. Two features are prominent:

(i) The bright transitions $\omega_{20}(X^0)$ and $\omega_{30}(X^0)$ are
polarized and the polarization anisotropy depends on height; see Fig.
\ref{Fig_4}(a).  Similarly, the lowest transitions of the biexciton
$\omega_{03}(XX^0)$ and $\omega_{02}(XX^0)$ are polarized and the degree of
polarization also depends on height. These transitions correspond to the {\em
  decay} into the two bright states of the monoexciton [Fig.  \ref{Fig_1}(d)].
Clearly, the biexciton transitions inherit the polarization of the monoexciton
transitions. For both $X^0$ and $XX^0$, we see that $\lambda$ switches sign as
a function of height.  In particular, at $h=50\,${\AA} the transitions of
$X^0$ present no in-plane polarization anisotropy [see arrow in Figs.
\ref{Fig_4}(a)] because the bright doublet formed by $|\Psi^{(2)}(X^0)\rangle$
and $|\Psi^{(3)}(X^0)\rangle$ is degenerate, which according to our
calculation results in $I^{([110])}(\omega_{20})\simeq I^{([1\bar
  10])}(\omega_{20})$.

(ii) The lowest optical transitions of $X^-$ and $X^+$ have degenerate
transition (recombination) energies [Fig. \ref{Fig_1}(b) and (c)] and,
according to our calculations, this results in $I^{([110])}(\omega_{00})\simeq
I^{([1\bar 10])}(\omega_{00})$ and, thus, negligible in-plane polarization
anisotropy regardless of height.

\begin{figure}[htb]
\includegraphics[width=8.5cm]{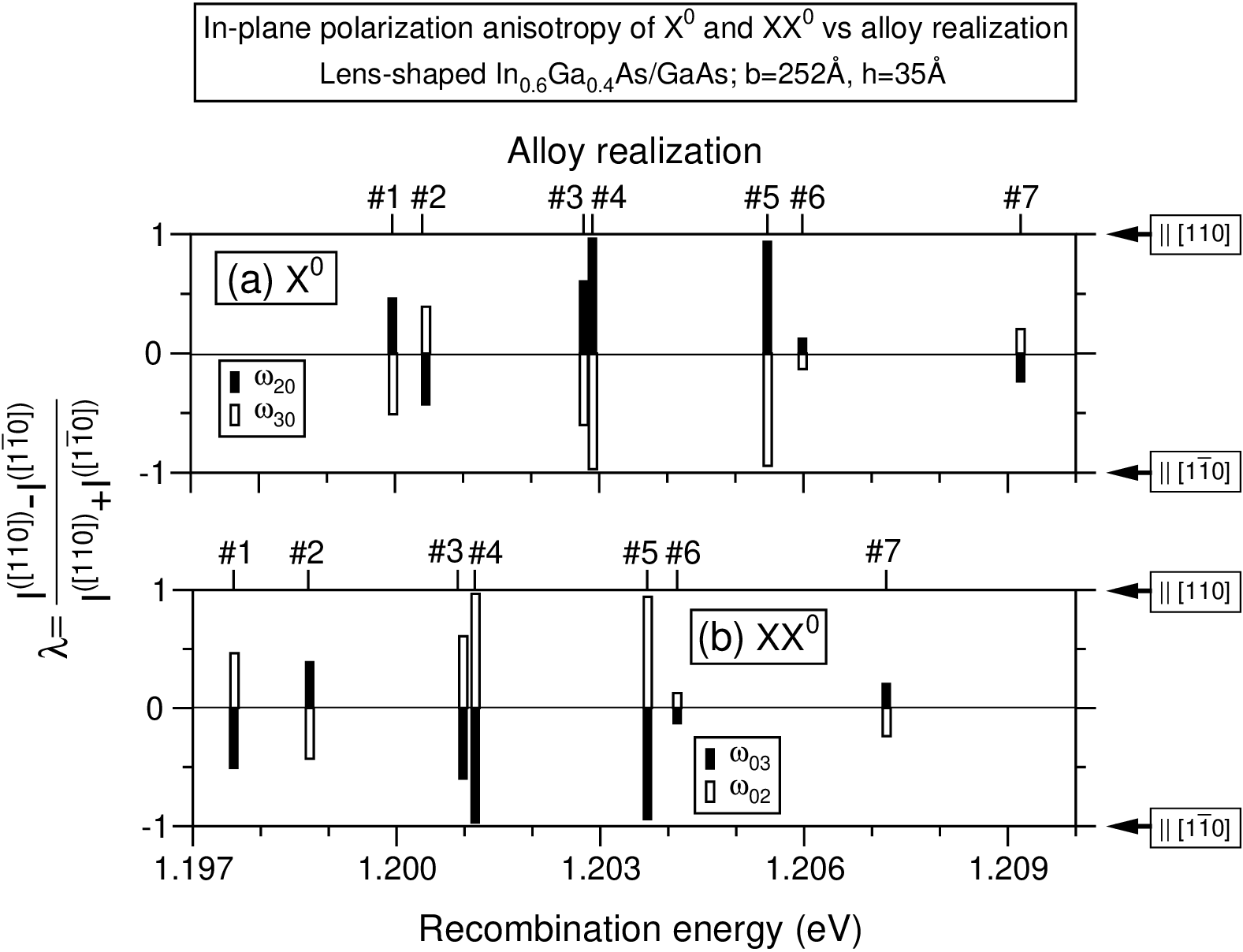}
\caption{{\label{Fig_5.new}}In-plane polarization $\lambda$ for the lowest
  optical transitions of (a) $X^0$ and (b) $XX^0$ for different realizations
  of alloy randomness in a $35\,${\AA}-tall In$_{\rm 0.6}$Ga$_{\rm
    0.4}$As/GaAs dot (base $b=252\,${\AA}). For both $X^0$ and $XX^0$,
  $\lambda$ depends dramatically on alloy realization, changing not only its
  sign but its magnitude too.}
\end{figure}

In addition to (i) and (ii), we find that (iii) the in-plane polarization
anisotropy of the lowest transitions of $X^0$ and $XX^0$ depends dramatically
on the dot's alloy randomness (disorder realization); as shown in Figure
\ref{Fig_5.new} for seven alloy realizations in a $35\,${\AA}-tall dot. For a
given transition in both $X^0$ and $XX^0$, $\lambda$ changes sign depending on
alloy randomness. Further, while some alloy realizations like $4$ and $5$
result in transitions nearly fully polarized ($\lambda\sim 99\%$), others such
as $2$, $6$ and $7$ present small anisotropy ($\lambda\sim 20\%$).

%

%
%
%
%
%
\section{Radiative recombination lifetimes}

The {\em characteristic} radiative lifetime $\tau_{if}(\chi^q)$ of a transition
$|\Psi^{(i)}(\chi^q)\rangle\rightarrow|\Psi^{(f)}(\chi^q-1)\rangle$ follows
from both the magnitude of the dipole matrix element of the transition
$\big|{\bf M}^{(\hat{\bf e})}_{if}(\chi^q)\big|^2$ and the recombination
energy $\omega_{if}$.\cite{dexter_book} Namely,

\begin{equation}
\label{intrinsic.lifetime}
\frac{1}{\tau_{if}(\chi^q)}=\frac{4}{3}\,\left(\frac{e^2}{m_0^2\,c^3\,\hbar^2}\right)n\,\omega_{if}(\chi^q)\sum_{\hat{{\bf
      e}}=\hat{x},\hat{y},\hat{z}}\big|{\bf M}^{(\hat{\bf e})}_{if}(\chi^q)\big|^2.
\end{equation}

\noindent Here, $e$ and $m_0$ are the charge and mass of the electron, 
respectively, and $c$ is the velocity of light in vacuum. 
In addition, the refractive index $n$ of the dot material accounts
for the material's effects on the photon emission. The linear dependence of
$1/\tau_{if}(\chi^q)$ on refractive index is applicable only when considering
dot and matrix materials with similar dielectric constants, as it is the case
in InGaAs/GaAs dots. In a more general case, more complicate dependences have
been proposed.\cite{thranhardt_PRB_2002} Note that the characteristic
radiative lifetime [Eq. (\ref{intrinsic.lifetime})] does not depend on
temperature nor on the occupation probability of the initial state
$|\Psi^{(i)}(\chi^q)\rangle$, as it is a characteristic property of the
transition $|\Psi^{(i)}(\chi^q)\rangle\rightarrow|\Psi^{(f)}(\chi^q-1)\rangle$. On the
other hand, the actual radiative lifetime $\tau(\chi^q)$ of exciton $\chi^q$
depends both on the probability $n_i$ of having the initial states
$|\Psi^{(i)}(\chi^q)\rangle$ of exciton $\chi^q$ occupied and the number of
final states $|\Psi^{(f)}(\chi^q-1)\rangle$ of exciton $\chi^q-1$ available
for recombination, as well as on the characteristic radiative lifetimes
$\tau_{if}(\chi^q)$. We calculate $\tau(\chi^q)$ from

\begin{equation}
\label{lifetime}
\frac{1}{\tau(\chi^q)}=\sum_f\sum_i\,n_i\frac{1}{\tau_{if}(\chi^q)}.
\end{equation} 

\noindent Here, $\sum_i\,n_i=1$ and $n_i=n_i(T;\chi^q)$ where $T$ is the 
temperature of the system. In general, the calculation of $n_i$ involves
solving a system of rate equations.\cite{dekel_PRB_2000} However, in the case
that the intra-level relaxation of exciton $\chi^q$ is much faster than the
radiative lifetimes, $n_i$ is given by the Boltzmann weight $P_i(T;\chi^q)$
[Eq.  (\ref{bolzmann.n})].

\subsection{Radiative lifetime $\tau(\chi^q)$ of the lowest optical
  transitions of $X^0$, $X^-$, $X^+$, and $XX^0$}

{\em Monoexciton.} At $T=0\,{\rm K}$, and assuming fast non-radiative
relaxation to the dark ground-state of $X^0$, the radiative lifetime of the
monoexciton $\tau(X^0)$ equals the characteristic radiative lifetime
$\tau_{00}(X^0)$ and it is long ($\sim {\rm ms}$). At finite temperatures, all
four levels [Fig. \ref{Fig_1}(a)] of the $X^0$ ground state are thermally
populated and emit light with their own characteristic lifetime [Eq.
(\ref{intrinsic.lifetime})].  For temperatures such that the lowest four
monoexciton states are occupied, while the occupation of excited states is
negligible, we calculate the radiative lifetime from Eq. (\ref{lifetime}) and
find

\begin{equation}
\label{tau.HT}
\tau(X^0)=4\,\left[\frac{\tau_{20}(X^0)\,\tau_{30}(X^0)}{\tau_{20}(X^0)+\tau_{30}(X^0)}\right]\simeq 2\,\tau_{20}(X^0).
\end{equation}

\begin{figure}[htb]
\includegraphics[width=8.5cm]{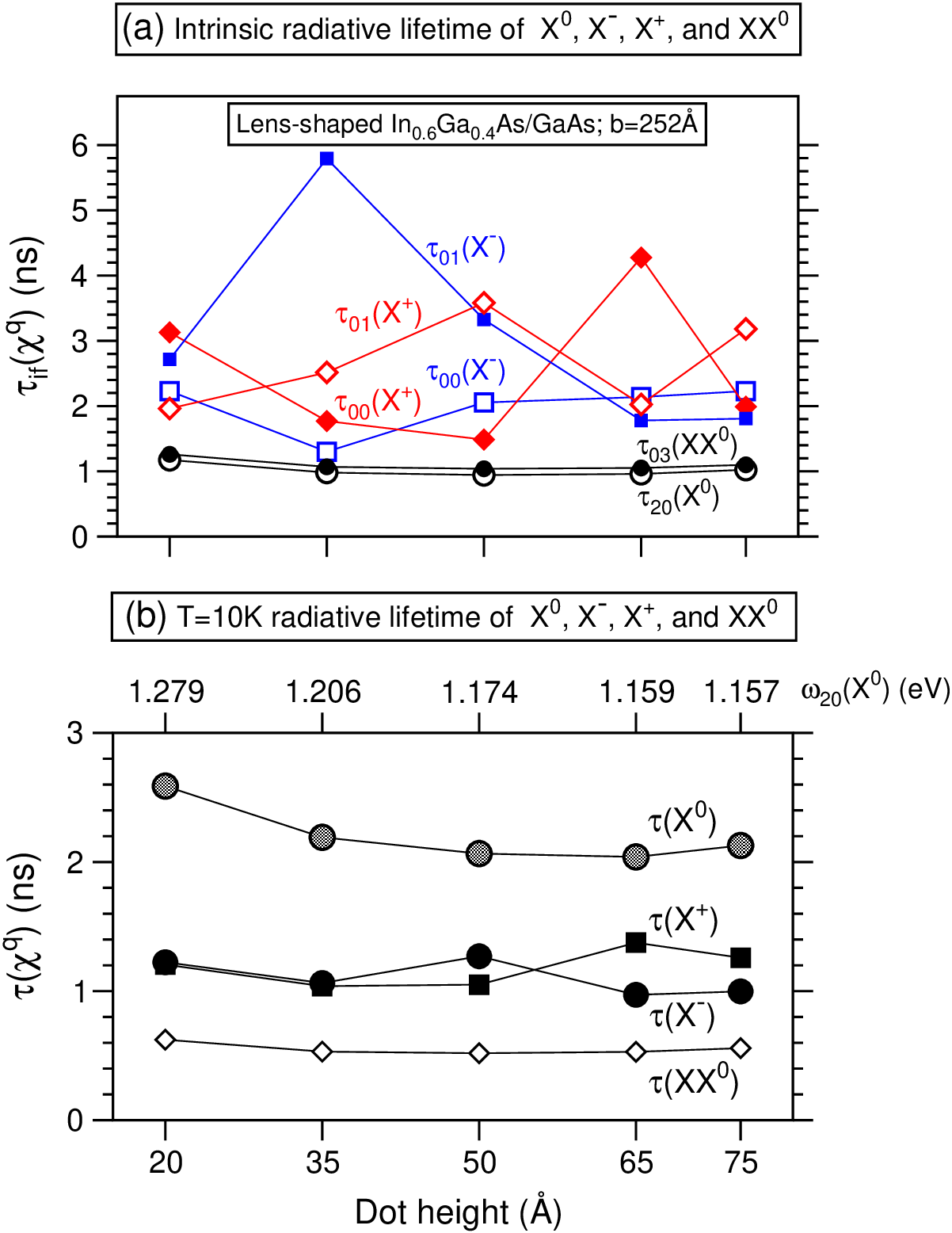}
\caption{{\label{Fig_5}}(Color online) (a) Characteristic radiative lifetime $\tau_{if}(\chi^q)$ 
  [Eq. (\ref{intrinsic.lifetime})] of the lowest optical transition of $X^0$,
  $X^-$, $X^+$, and $XX^0$ as a function of dot height. For $X^0$ and $XX^0$,
  the characteristic lifetimes are nearly the same and depend weakly on
  height. A stronger dependence on height is found for $X^-$ and $X^+$. (b)
  $T=10\,{\rm K}$ radiative lifetime $\tau(\chi^q)$ [Eq. (\ref{lifetime})] of
  the lowest optical transition of $X^0$, $X^-$, $X^+$, and $XX^0$ as a
  function of dot height.  The lifetimes depend weakly on height for all these
  excitons. For $X^0$ and $XX^0$, the lifetimes are nearly the same. A similar
  trend is found for $X^-$ and $X^+$.  At all heights, the radiative lifetimes
  of $X^-$ and $X^+$ are about twice as big as that of $XX^0$.}  
\end{figure}

\noindent Here, we have used a result of our calculations that predict
$\tau_{20}(X^0)\simeq\tau_{30}(X^0)$ regardless of dot height. In addition, we
have neglected the long-lived recombinations $\omega_{00}$ and $\omega_{10}$.
Figure \ref{Fig_5}(a) shows the characteristic $\tau_{20}(X^0)$ versus dot
height.  We find that this lifetime depends weakly on height. A calculation of
$\tau(X^0)$ at $T=10\,{\rm K}$ predicts a moderate decrease in the monoexciton
radiative lifetime as height increases [Fig.  \ref{Fig_5}(b)]. It should be
noted that $\tau(X^0)$ is actually bigger than the approximate value of
$2\tau_{20}(X^0)$ [Eq. (\ref{tau.HT})] due to the actual occupation
probability $n_i(10\,{\rm K},X^0)$ of the initial states. By measuring
time-resolved photoluminescence in InAs/GaAs dots ($b=120$-$220\;${\AA},
height unspecified) with monoexciton emission ranging from $1.2$-$1.3\;{\rm
  eV}$, Karachinsky and coworkers have recently observed the opposite
trend,\cite{karachinsky_APL_2004} i. e. $\tau(X^0)$ increases with dot size
from $1\;{\rm ns}$ in a dot with emission energy of $1.31\;{\rm eV}$ to $\sim
3\;{\rm ns}$ in one with emission energy of $1.25\;{\rm eV}$. The values of
$\tau(X^0)$ we predict for tall dots ($h=65\;${\AA} and $75\;${\AA}) agree
satisfactorily with the value of $1\;{\rm ns}$ measured by Buckle {\em et al}
for an InAs/GaAs dot ($b\sim 120\;${\AA}, $h\sim\;30${\AA}) with gap
$1.131\;{\rm eV}$ (Ref.  \onlinecite{buckle_JAP_1999}) at a temperature of
$6\;{\rm K}$. Further, our predictions also agree well with the value of
$1.55\;{\rm ns}$ extracted from time-resolved photoluminescence experiments at
$10\,{\rm K}$ performed in InAs/GaAs dots ($b=200\;{\AA}$, $h=20\;${\AA}) by
Bardot and coworkers.\cite{bardot_condmat_2005}

%

%
%
{\em Biexciton.} In contrast to the monoexciton, the biexciton ground state
is singly-degenerate and bright [Fig. \ref{Fig_1}(b)]. Thus, at $T=0\,{\rm
  K}$, the radiative lifetime is given from Eq. (\ref{lifetime}) as

\begin{equation}
\label{tau.XX0}
\tau(XX^0)=\left[\frac{\tau_{02}(XX^0)\,\tau_{03}(XX^0)}{\tau_{02}(XX^0)+\tau_{03}(XX^0)}\right]\simeq \frac{1}{2}\,\tau_{03}(XX^0).
\end{equation}

\noindent Similarly to the $X^0$ case, in Eq. (\ref{tau.XX0}) we have used the
relationship $\tau_{03}(XX^0)\simeq\tau_{02}(XX^0)$ that our calculations
predict, and we have neglected the long-lived dark recombination channels.
Figure \ref{Fig_5}(a) shows $\tau_{03}(XX^0)$ as a function of height.  We
find a weak dependence with height, as in the monoexciton case.  Remarkably,
we find that $\tau_{20}(X^0)\simeq\tau_{03}(XX^0)$ regardless of height. The
latter leads to the following relationship between the radiative lifetime of the
monoexciton and biexciton,

\begin{equation}
\tau(XX^0)\simeq\frac{1}{4}\,\tau(X^0).
\end{equation}

Figure \ref{Fig_5}(b) shows that at $T=10\,{\rm K}$ the calculated
$\tau(XX^0)$ depends weakly on height, changing by about $0.1\,{\rm ns}$ in
the entire range of heights studied. $\tau(XX^0)$ is in excellent agreement
with the value of $0.5\;{\rm ns}$ measured by Ulrich {\em et al} in an
(In,Ga)As/GaAs quantum dot ($b=150$-$200\;${\AA}, $h=10$-$20\;${\AA}) with an
exciton gap of $1.337\,{\rm eV}$ (Ref.  \onlinecite{ulrich_condmat_2004}).

%
%
{\em Trions.} Both $X^-$ and $X^+$ have a two-fold degenerate ground state
that is bright [Figs. \ref{Fig_1}(c) and (d)]. In each of these excitons,
there are four lowest transitions; namely, $\omega_{00}(X^-)$ [Fig.
\ref{Fig_1}(c)], $\omega_{10}(X^-)$, $\omega_{01}(X^-)$, and
$\omega_{11}(X^-)$; and $\omega_{00}(X^+)$ [Fig. \ref{Fig_1}(d)],
$\omega_{10}(X^+)$, $\omega_{01}(X^+)$, and $\omega_{11}(X^+)$, for $X^-$ and
$X^+$, respectively. In turn, each of these transitions have a corresponding
characteristic radiative lifetime. For the latter, our calculations predict
$\tau_{00}(X^-)=\tau_{11}(X^-)$ and $\tau_{01}(X^-)=\tau_{10}(X^-)$ as well as
$\tau_{00}(X^+)=\tau_{11}(X^+)$ and $\tau_{01}(X^+)=\tau_{10}(X^+)$.
Thus, at $T=0\,{\rm K}$, the radiative lifetimes of $X^-$ and $X^+$ are given by

\begin{widetext}
\begin{eqnarray}
\tau(X^-)=\left\{\frac{1}{2}\left[\frac{1}{\tau_{00}(X^-)}+\frac{1}{\tau_{01}(X^-)}\right]+\frac{1}{2}\left[\frac{1}{\tau_{10}(X^-)}+\frac{1}{\tau_{11}(X^-)}\right]\right\}^{-1}=\frac{\tau_{00}(X^-)\tau_{01}(X^-)}{\tau_{00}(X^-)+\tau_{01}(X^-)},\nonumber\\
\\
\tau(X^+)=\left\{\frac{1}{2}\left[\frac{1}{\tau_{00}(X^+)}+\frac{1}{\tau_{01}(X^+)}\right]+\frac{1}{2}\left[\frac{1}{\tau_{10}(X^+)}+\frac{1}{\tau_{11}(X^+)}\right]\right\}^{-1}=\frac{\tau_{00}(X^+)\tau_{01}(X^+)}{\tau_{00}(X^+)+\tau_{01}(X^+)}.\nonumber\\
\end{eqnarray}
\end{widetext}

\noindent Figure \ref{Fig_5}(a) shows the height dependence of $\tau_{00}(X^-)$
and $\tau_{01}(X^-)$, and $\tau_{00}(X^+)$ and $\tau_{01}(X^+)$. These {\em
  characteristic} radiative lifetimes depend strongly and non-monotonically on
height. This non-monotonic dependence translates into a rather simple and
monotonic $\tau(X^-)$ and $\tau(X^+)$, as shown in Fig. \ref{Fig_5}(b). For
flat dots $\tau(X^-)\simeq\tau(X^+)$, whereas for taller dots these lifetimes
become slightly different. Our predicted $\tau(X^-)$ are in satisfactory
agreement with the value of $0.6\;{\rm ns}$ recently observed by Smith and
co-workers in (In,Ga)As/GaAs dots (size unspecified) with exciton ground-state
emission at $\sim 1.318\,{\rm eV}$ (Ref. \onlinecite{smith_PRL_2005}). We find
that the radiative lifetimes of the charged trions satisfy the relationship
$\tau(X^0)>\tau(X^-)\sim\tau(X^+)>\tau(XX^0)$.

\section{Comparison of $X^0$, $X^-$, $X^+$, and $XX^0$ in 
lens-shaped pure InAs/GaAs with alloyed (In,Ga)As/GaAs dots} 

For completeness, we briefly compare the binding and recombination energies,
polarization anisotropy, and radiative lifetimes in lens-shaped pure,
non-alloyed InAs/GaAs dots with In$_{0.6}$Ga$_{0.4}$As/GaAs dots.  Note that
Williamson, Wang, and Zunger have already compared results for $X^0$ for
several alloy profiles.\cite{williamson_PRB_2000} In Ref.
\onlinecite{narvaez_condmat} we predicted that hole localization takes place
at the dot-GaAs matrix interface as the height of these dots increases above
$35\;${\AA}.  Thus, we discuss here two flat dots ($h=20\;${\AA} and
$35\;${\AA}) with base $b=252\;${\AA}.

(i) Recombination energies are smaller in InAs/GaAs dots than in In$_{\rm
  0.6}$Ga$_{\rm 0.4}$As/GaAs dots with the same geometry. For instance,
$\omega_{20}(X^0)=1.078\;{\rm eV}$ and $0.987\;{\rm eV}$ for $h=20\;${\AA} and
$35\;${\AA}, respectively.

(ii) The spectroscopic shifts show the same trends with height in flat
InAs/GaAs as those in flat In$_{\rm 0.6}$Ga$_{\rm 0.4}$As/GaAs dots. So do the
binding energies $\Delta(X^-)$ and $\Delta(X^+)$, and $\Delta(XX^0)$. However,
there are two important differences bewteen the pure, non-alloyed InAs/GaAs
dots and their In$_{\rm 0.6}$Ga$_{\rm 0.4}$As/GaAs counterparts: (a) For the
$20\;$\AA-tall InAs/GaAs dot, while the binding energies still satisfy
$\Delta(X^-) > \Delta(XX^0) > \Delta(X^+)$, we find that $X^+$ and $XX^0$ are
{\em bound}, with $\Delta(X^+)=1.6\;{\rm meV}$ and $\Delta(XX^0)=1.5\;{\rm
  meV}$, respectively. This is so because at this height the InAs/GaAs dot is
in the nearly ``symmetric'' regime: $J^{(hh)}_{00}=25.6\;{\rm meV}\sim
J^{(ee)}_{00}=25.1\;{\rm meV} \sim J^{(eh)}_{00}=25.3\;{\rm meV}$, so the
Hartree-Fock component of the binding energy [see Eq. (\ref{Correlation}) for
$XX^0$ case] is much smaller than in the In$_{\rm 0.6}$Ga$_{0.4}$As/GaAs dot
with same height.  Thus, correlation becomes capable of binding $X^+$ and
$XX^0$ in the pure, non-alloyed dot.  (b) For the $35\;$\AA-tall InAs/GaAs
dot, we find {\em ordering reversal}, i.e.  $\Delta(X^+) > \Delta(XX^0) >
\Delta(X^-)$. In In$_{0.6}$Ga$_{0.4}$As/GaAs dots, this ordering is attained
at $h=50\;${\AA} (Fig. \ref{Fig_3.mp}).

(iii) In contrast to the findings in In$_{\rm 0.6}$Ga$_{\rm 0.4}$As/GaAs dots,
transitions $\omega_{20}(X^0)$ and $\omega_{30}(X^0)$ are fully polarized
along $[110]$ ($\lambda=1$) and $[1\bar 10]$ ($\lambda=-1$), respectively,
regardless of height. Consequently, $\omega_{03}(XX^0)$ and
$\omega_{02}(XX^0)$ are fully polarized along $[1\bar 10]$ and $[110]$,
respectively. These polarizations are expected from a dot with $C_{2v}$
symmetry, like a lens-shaped pure InAs/GaAs dot.

(iv) Radiative lifetimes $\tau(X^0)$, $\tau(X^+)$ and $\tau(X^-)$, and
$\tau(XX^0)$ are similar to those in In$_{\rm 0.6}$Ga$_{\rm
  0.4}$As/GaAs dots with the same geometry. For instance, $\tau(X^0)=2.8\;{\rm
  ns}$ and $\tau(XX^0)=0.7\;{\rm ns}$ for $h=35\;${\AA}; and
$\tau(X^0)=2.9\;{\rm ns}$ and $\tau(XX^0)=0.6\;{\rm ns}$ for $h=20\;${\AA}.

%
%
%
%
%
\section{Summary}

We have addressed the height dependence of recombination energies,
polarization and radiative lifetimes of the lowest optical transitions of the
neutral exciton ($X^0$), negatively- ($X^{-}$) and positively-charged
($X^{+}$) trions, and the biexciton ($XX^0$) in lens-shaped, self assembled
In$_{0.6}$Ga$_{0.4}$As/GaAs quantum dots. We have predicted the following.

(i) The recombination energy of the lowest transition of $X^-$, $X^+$ and
$XX^0$, correspondingly, $\omega_{00}(X^-)$, $\omega_{00}(X^+)$ and
$\omega_{03}(XX^0)$ shows qualitatively different behavior for each excitonic
complex.  Namely, $\omega_{00}(X^-)$ blue-shifts as height increases, whereas
that of $\omega_{00}(X^+)$ red-shifts. On the other hand, as height increases,
$\omega_{03}(XX^0)$ shows a red-shift at small heights, reaches a maximum
shift, and then blue-shifts for taller dots. This behavior is explained by the
height dependence and relative magnitude of $J^{\,(ee)}_{00}$,
$J^{\,(hh)}_{00}$ and $J^{\,(eh)}_{00}$.

(ii) The binding energies $\Delta(X^-)$, $\Delta(X^+)$ and $\Delta(XX^0)$
follow the height dependence of the emission spectroscopic shifts. Changes in
the dot height drives a bound-to-unbound crossover for each of these
complexes.

(iii) The in-plane polarization anisotropy $\lambda$ of the lowest transitions
of $X^0$ ($\omega_{20}$) and $XX^0$ ($\omega_{03}$) strongly depends on dot
height as well as on alloy randomness (disorder realization).  In contrast,
the lowest transitions of $X^-$ and $X^+$ present negligible $\lambda$
regardless of height.

(iv) The ground state of $X^0$ encompasses four states that split off in a
low-energy pair that is dark and a high-energy pair that is bright, with a
bright-dark splitting that increases as height decreases.  Thus, at $T=0\;{\rm
  K}$ the radiative lifetime $\tau(X^0)$ of $X^0$ is long.  On the other hand,
at $T=10\;{\rm K}$ both dark and bright states are populated; so, $\tau(X^0)$
becomes fast, moderately decreases as height increases, and its magnitude
ranges from $2$-$3\;{\rm ns}$.  In contrast, $\tau(X^-)$, $\tau(X^+)$ and
$\tau(XX^0)$ are fast even at $T=0\;{\rm K}$, as a consequence of these
excitons having ground states that are bright. These radiative lifetimes
depend weakly on height. Further, $\tau(X^-)\sim\tau(X^+)\simeq 1.1\;{\rm
  ns}$, while $\tau(XX^0)\simeq 0.5\;{\rm ns}$.

We have compared these predictions with available data and have found them in
satisfactory agreement. In addition, we compared with results in pure,
non-alloyed InGa/GaAs quantum dots.

%
%
%
\begin{acknowledgments}
  We thank Alberto Franceschetti (NREL) for valuable discussions. This work
  has been supported by U.S. DOE-SC-BES-DMS under contract No.
  DE-AC36-99GO10337.
\end{acknowledgments}

%

%

\end{document}